\documentclass[]{emulateapj}
\usepackage{natbib,amsmath}

\begin{document}

\def\gq{$\Gamma_{\rm q}$}
\def\mgq{\Gamma_{\rm q}}
\def\meq{\epsilon^{\rm q}_{912}}
\def\mgigm{\Gamma_{\rm IGM}}
\def\gigm{$\Gamma_{\rm IGM}$}
\def\mug{(u-g)}
\def\ug{$(u-g)$}
\def\nbin{150}
\def\vbestmfp{48.4 \pm 2.1}
\def\vslopemfp{38.0 \pm 5.3}
\def\bestmfp{\lambda_0}
\def\slopemfp{b_\lambda}
\def\omegam{\Omega_{\rm m}}
\def\mnull{\nu_{\rm 912}}
\def\nnull{$\nu_{\rm 912}$}
\def\hub{h_{72}^{-1}}
\def\umfp{{\hub \, \rm Mpc}}
\def\lbr{$\lambda_{\rm r}$}
\def\mlbr{\lambda_{\rm r}}
\def\intl{\int\limits}
\def\kll{$\kappa_{\rm LL}$}
\def\mkll{\kappa_{\rm LL}}
\def\kconst{$\kappa_{\mzq}$}
\def\mkconst{\kappa_{\mzq}}
\def\mztkll{{\tilde\kappa}_{912}}
\def\zq{$z_q$}
\def\mzq{z_q}
\def\maxoff{0.4}
\def\clls{3.50 \pm 0.12}
\def\alls{1.89 \pm 0.39}
\def\blls{-1.0 \pm 0.3}
\def\cmma{\;\;\; ,}
\def\perd{\;\;\; .}
\def\ltk{\left [ \,}
\def\ltp{\left ( \,}
\def\ltb{\left \{ \,}
\def\rtk{\, \right  ] }
\def\rtp{\, \right  ) }
\def\rtb{\, \right \} }
\def\sci#1{{\; \times \; 10^{#1}}}
\def \rAA {\rm \AA}
\def \zem {$z_{\rm em}$}
\def \mzem {z_{\rm em}}
\def \mzlls {z_{\rm LLS}}
\def \zlls {$z_{\rm LLS}$}
\def \mzend {z_{\rm end}}
\def \zend {$z_{\rm end}$}
\def \zstrto {$z_{\rm start}^{\rm S/N=1}$}
\def \mzstrto {z_{\rm start}^{\rm S/N=1}}
\def \zstrt {$z_{\rm start}$}
\def \mzstrt {z_{\rm start}}
\def\smm{\sum\limits}
\def \lll  {$\lambda_{\rm 912}$}
\def \mlll  {\lambda_{\rm 912}}
\def \mtll  {\tau_{\rm 912}}
\def \tll  {$\tau_{\rm 912}$}
\def \avtll  {\tilde{\mtll}}
\def \tigm  {$\tau_{\rm IGM}$}
\def \mtigm  {\tau_{\rm IGM}}
\def \cmm  {cm$^{-2}$}
\def \cmmm {cm$^{-3}$}
\def \kms  {km~s$^{-1}$}
\def \mkms  {{\rm km~s^{-1}}}
\def \lyaf {Ly$\alpha$ forest}
\def \Lya  {Ly$\alpha$}
\def \lya  {Ly$\alpha$}
\def \mlya  {Ly\alpha}
\def \Lyb  {Ly$\beta$}
\def \lyb  {Ly$\beta$}
\def \Lyg  {Ly$\gamma$}
\def \lyg  {Ly$\gamma$}
\def \lyd  {Ly$\delta$}
\def \ly5  {Ly-5}
\def \ly6  {Ly-6}
\def \ly7  {Ly-7}
\def \nhi  {$N_{\rm HI}$}
\def \mnhi  {N_{\rm HI}}
\def \lnhi {$\log N_{HI}$}
\def \mlnhi {\log N_{HI}}
\def \etal {\textit{et al.}}
\def \ob {$\Omega_b$}
\def \obh {$\Omega_bh^{-2}$}
\def \om {$\Omega_m$}
\def \ol {$\Omega_{\Lambda}$}
\def \gz {$g(z)$}
\def \mgz {g(z)}
\def \lyaf {Lyman--$\alpha$ forest}
\def \fnhi {$f(\mnhi,z)$}
\def \mfnhi {f(\mnhi,z)}
\def \mfp {$\lambda_{\rm mfp}^{912}$}
\def \mmfp {\lambda_{\rm mfp}^{912}}
\def \btlls {$\beta_{\rm LLS}$}
\def \mbtlls {\beta_{\rm LLS}}
\def \teff {$\tau_{\rm eff,LL}$}
\def \mteff {\tau_{\rm eff,LL}}
\def \O {${\mathcal O}(N,X)$}
\newcommand{\cm}[1]{\, {\rm cm^{#1}}}
\def \zem {$z_{em}$}
\def \snrlim {SNR$_{lim}$}

\title{A Direct Measurement of the IGM 
Opacity to \ion{H}{1} Ionizing Photons}

\author{
J. Xavier Prochaska\altaffilmark{1}, 
Gabor Worseck\altaffilmark{1}, 
John M. O'Meara\altaffilmark{2}, 
}
\altaffiltext{1}{Department of Astronomy and Astrophysics, UCO/Lick Observatory, University of California, 1156 High Street, Santa Cruz, CA 95064}
\altaffiltext{2}{Department of Chemistry and Physics, Saint Michael's College.
One Winooski Park, Colchester, VT 05439}

\begin{abstract}
We present a new method to directly measure the opacity from
\ion{H}{1} Lyman limit (LL) absorption \kll\ along quasar sightlines
by the intergalactic medium (IGM).  
The approach analyzes the average (``stacked'') spectrum of an ensemble
of quasars at a common redshift to infer the mean free path
\mfp\ to ionizing radiation.  We apply this technique to 
1800 quasars at $z = 3.50 - 4.34$ drawn from the Sloan Digital
Sky Survey (SDSS), giving the most precise measurements on \kll\
at any redshift.
From $z=3.6$ to 4.3, the opacity increases steadily 
as expected and is well parameterized by
$\mmfp = \bestmfp - \slopemfp (z-3.6)$ with 
$\bestmfp = (\vbestmfp) \, \umfp$ 
and $\slopemfp = (\vslopemfp) \, \umfp$ (proper distance).
The relatively high \mfp\ values indicate that the incidence
of systems which dominate 
\kll\ evolves less strongly at $z>3$ than that of the \lya\ forest. 
We infer a mean free path three times higher than some previous 
estimates, a result which has important implications for the 
photo-ionization rate derived from the emissivity of star 
forming galaxies and quasars.
Finally, our analysis reveals a previously unreported,
systematic bias in the SDSS quasar sample
related to the survey's color targeting criteria.
This bias potentially affects all $z \sim 3$
IGM studies using the SDSS database.
\end{abstract}


\section{Introduction}

The observed high transmission of $z \sim 3$ quasars at rest
wavelengths \lbr\ blueward of \ion{H}{1} \lya\ reveals that the 
intergalactic medium (IGM) is highly ionized \citep{gp65}.
The presence of the \lya\ forest demands an intense, extragalactic
ultraviolet background (EUVB) radiation field. 
The quasars themselves provide a significant fraction of the
required ionizing flux, buoyed
by the emission from more numerous yet fainter star-forming galaxies.
Several recent studies have argued that the latter population
dominates the EUVB at $z \gtrsim 3$ \citep{flh+08,cbt09,dww09},
where the quasar population likely declines \citep[e.g.][]{fan04}.
These assertions, however, hinge on the opacity of the
IGM to ionizing radiation via \ion{H}{1} Lyman limit absorption (\kll)
which directly impacts estimates of the EUVB
measured from the integrated quasar and stellar ionizing emissivity.

Traditionally, \kll\ has been estimated from the incidence of
so-called Lyman limit systems (LLSs) via surveys of quasar
spectroscopy \citep[e.g.][]{lzt91,storrie94,peroux_dla03}.
Observationally, one can rather easily identify systems with
large optical depths $\mtll \gtrsim 2$ and
the majority of these surveys have probed to this limit.
The integrated opacity of the IGM, however, includes and
is likely dominated by gas with $\mtll < 1$, the so-called
partial Lyman limit systems and \lya\ forest clouds.
Systems with these \ion{H}{1} column densities
($\mnhi \approx 10^{14} - 10^{17} \cm{-2}$) are especially
difficult to survey because the strong lines of the Lyman series
(e.g.\ \lya, \lyb)
lie on the flat portion of the curve-of-growth and they
exhibit only weak absorption at the Lyman limit.
Therefore, current estimates of \kll\ are based on an 
extrapolation/interpolation of the frequency of systems 
with $\mnhi < 10^{14} \cm{-2}$ and $\mnhi > 10^{17.5} \cm{-2}$
\citep{mhr99,sb03,flz+09}.
Current constraints on \kll\ span over a magnitude of uncertainty,
especially at $z \gtrsim 4$.


In this Letter, we introduce a new technique to estimate \kll\ 
that avoids the traditional line-counting statistics of the IGM.
We analyze the average rest-frame spectra of  
1800 $z>3.5$ quasars drawn from the Sloan Digital Sky Survey,
Data Release 7 \citep{sdssdr7}.
As an ensemble at a common redshift, 
these ``stacked'' spectra show the exponential
drop in flux at $\lambda < \mlll = c/\mnull = 911.76$\AA\ from the integrated
opacity of the IGM.  We precisely evaluate \kll\ in a series of
small redshift intervals covering $z \approx 3.6 - 4.3$
to explore redshift evolution.  
We adopt a cosmology with 
$H_0 = 72 \, h_{72} \, \mkms \, \rm Mpc^{-1}$, 
$\Omega_{\rm m} = 0.3$, and $\Omega_\Lambda = 0.7$
and report proper lengths unless specified.

\begin{figure*}
\begin{center}
\includegraphics[height=6.8in,angle=90]{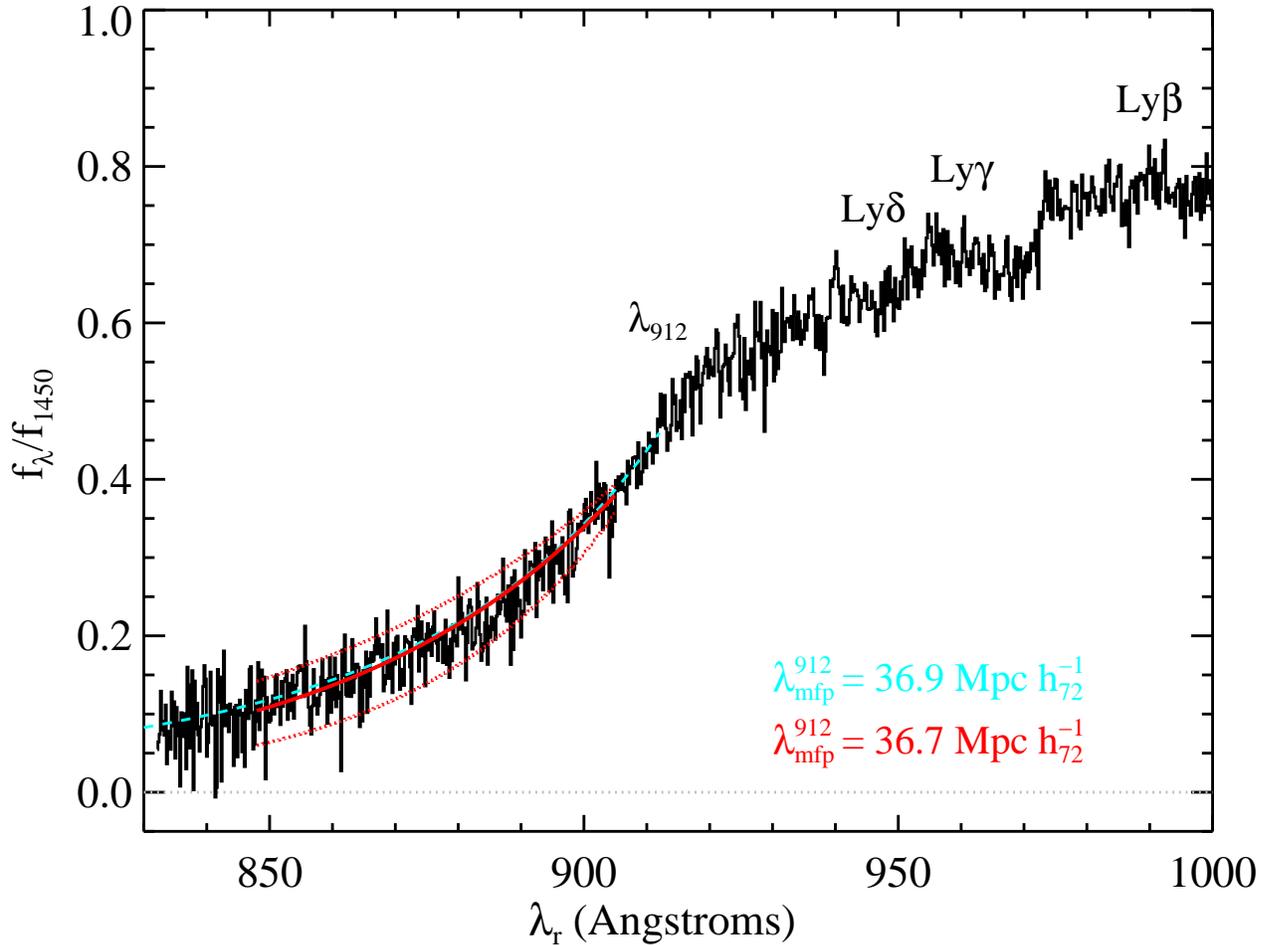}
\end{center}
\caption{The stacked spectrum (normalized at $\mlbr = 1450$\AA)
from 150 mock quasar spectra with IGM absorption derived
from an assumed \fnhi\ distribution.  
The adopted emission redshifts and S/N of the spectra were
taken to match our SDSS quasar sample at $z=[3.59,3.63]$.
The sharp decline in the flux at
$\mlbr \approx 912$\AA\ is associated first with the higher order
Lyman series opacity (for $\mlbr \gtrsim 912$\AA) and then
Lyman limit continuum opacity (for $\mlbr < 912$\AA).
The dashed (cyan) curve overplotted on the spectrum is the
predicted relative flux corresponding to \teff\ 
(Equation~\ref{eqn:teff}), normalized at $\mlbr = 912$\AA.
On top of this curve is the best-fit model of the flux
evaluated over the
interval $\mlbr = 848-905$\AA\ using our new methodology
(Equation~\ref{eqn:newteff}).
The resulting estimates for the mean free path \mfp\ 
are indicated on the figure and are in excellent agreement.
Finally, the dotted curves show the predicted
models after offsetting \mfp\ by $\pm 10 \, \rm Mpc$ and
indicate the sensitivity of our technique to measuring \mfp.
}
\label{fig:mock}
\end{figure*}

\section{Methodology}

The \ion{H}{1} Lyman limit opacity of the IGM has traditionally
been expressed as an effective optical depth \teff\ estimated
from the observationally constrained \nhi\ frequency distribution 
\fnhi.  An ionizing photon ($\nu \ge \mnull$) emitted from 
a quasar with redshift
$z = \mzq$ will redshift to 1\,Ryd 
at $z = z_{912} \equiv (\nu_{\rm 912}/\nu)(1+\mzq) -1 $.
The effective optical depth that this photon experiences 
by IGM Lyman limit opacity is then \citep[cf.][]{mm93}:
\begin{eqnarray}
&& \mteff(z_{912},\mzq) = \label{eqn:teff}\\
&& \intl_{z_{912}}^{z_q} \intl_0^\infty f(\mnhi,z')
   \lbrace 1 - \exp \ltk - \mnhi \sigma_{\rm ph}(z') \rtk \rbrace d\mnhi dz'  \nonumber
\end{eqnarray}
where $\sigma_{\rm ph}$ is the photoionization cross-section evaluated
at the photon frequency.   In practice, this approach is subject to
large uncertainties because 
(1) \fnhi\ is poorly constrained 
for systems with $\mtll \lesssim 1$,
(2) observational surveys rarely measure \fnhi\ at the same epoch forcing
interpolation and extrapolation; 
and
(3) estimates of \fnhi\ for the LLS may always be subject to large
systematic uncertainty \citep{pow09}.
This traditional approach may never 
yield a precise and robust estimate of \teff. 

Our new approach is to directly measure \teff\ through analysis of
averaged ensembles of quasar spectra. 
In Figure~\ref{fig:mock}, we present the stacked spectrum of
\nbin\ mock quasar spectra at $z \approx 3.6$.
Each spectrum was given a unique emission
redshift \zq\ and SED\footnote{Quasar SEDs do not
show strong emission features at $\mlbr < 912$\AA\ \citep{telfer02}.} 
(normalized at 1450\AA), 
and then was blanketed with Lyman series and Lyman limit
absorption from an assumed \fnhi\ distribution 
\citep{dww08,wp09}.
The spectra were degraded to the nominal spectral resolution of the
SDSS spectrometer (FWHM=150\kms)
and Gaussian noise was added to give
a distribution\footnote{The S/N and emission
redshift distributions correspond to the third bin in the observational
analysis that follows.}  
of S/N values at $\mlbr = 1450$\AA. 
The data was then averaged without weighting.

Inspecting Figure~\ref{fig:mock}, one identifies the effective opacity
of the \lyb\ forest at $\mlbr \approx 1000$\AA\ and corresponding decrements
in the spectrum at \lyg\ and \lyd.  One then observes a steep drop
in the flux starting at $\mlbr \approx 920$\AA\ due 
to the opacity of higher order
Lyman series lines of optically thick absorbers (e.g.\ damped \lya\ systems).
The continued decline at $\mlbr < \mlll$, however, is dominated by the
continuum opacity of \ion{H}{1}.
At all wavelengths, the scatter in the stacked spectrum is due
to small-scale variance in IGM absorption, 
not noise in the individual spectra.  

Overplotted in Figure~\ref{fig:mock} 
is the flux model $f = f_{\rm 912} \exp[-\mteff(z)]$ with \teff\ evaluated
from the input \fnhi\ distribution (Equation~\ref{eqn:teff}) and
$f_{\rm 912}$, the flux at $\lambda = \mlll$, estimated from the data.
This is a good model of the stacked spectrum;
even though the underlying average SED evolves as 
$f_\lambda \propto \lambda^{2.4}$,
the analysis is performed over too small a wavelength interval
to note its evolution.
The evaluation gives $\mteff = 1$ at $(\mzq-z_{912})=0.22$ corresponding to 
a proper mean free path $\mmfp = 36.9 \, \umfp$ at $z=3.6$.

Now consider an alternate evaluation of \teff\ which follows the
standard definition of optical depth,

\begin{equation}
\mteff(r,\nu) = \intl_0^r \mkll(r',\nu) \, dr' \cmma
\label{eqn:odepth}
\end{equation}
where the integral is evaluated to an arbitrary proper distance from the 
quasar.  In an expanding universe, an ionizing photon emitted by
the quasar will be attenuated by the Lyman limit opacity \kll\ 
until it is redshifted to $h\nu = 1$\,Ryd at $z=z_{912}$.
If the photon is not absorbed by IGM line opacity from gas at $z<z_{912}$, 
it may be observed
today at a wavelength $\lambda_{\rm obs} = (1+z_{912}) c / \nu_{\rm 912}$.
During the photon's travel from \zq\ to $z_{912}$, the opacity \kll\ evolves
because of the decreasing frequency (redshift) and also from changes
to the physical conditions of the universe 
(e.g.\ the expanding proper distance).   We separate 
the frequency and radial dependencies in the opacity as follows, 

\begin{equation}
\mkll(r,\nu) \equiv \mztkll(r) \ltp \frac{\nu}{\mnull} \rtp^{-3}.
\label{eqn:kz}
\end{equation}
where the frequency dependence related to $\sigma_{\rm ph}$ is approximate.
This treatment also ignores stimulated emission, i.e.\ it assumes
\teff\ is dominated by `clouds' with $\mtll \lesssim 1$.

In principle, one can adopt
any radial dependence for \kll.  Expressing Equation~\ref{eqn:kz} 
in redshift space, we have

\begin{equation}
\mkll(z) = \mztkll(z) \ltp \frac{1+z}{1+z_{912}} \rtp^{-3}.
\end{equation}
With this functional form for the Lyman limit opacity, it is straightforward
to integrate Equation~\ref{eqn:odepth} by adopting a Friedman-Walker cosmology
where 

\begin{equation}
\frac{dr}{dz} \equiv \frac{c}{H(z) (1+z)} 
  = \frac{c/H_0}{(1+z) \sqrt{\Omega_{\rm m} (1+z)^3 + \Omega_\Lambda}}.
\end{equation}
At $z>3$, the universe is matter dominated and we can express
$dr/dz \approx c/(H_0 \Omega_{\rm m}^{1/2}) (1+z)^{-5/2}$.
Altogether, we find

\begin{eqnarray}
&& \mteff(z_{912},\mzq) = \nonumber \\
&& \frac{c}{H_0 \omegam^{1/2}} (1+z_{912})^3 
\intl_{z_{912}}^{\mzq} \mztkll(z') \, (1+z')^{-11/2} \, dz'.
\label{eqn:newteff}
\end{eqnarray}

In practice, we find that the analysis of a single stacked spectrum does
not constrain the redshift evolution in \kll.  
Therefore, we have simply
parameterized \kll\ by its value at $z=\mzq$,  i.e., $\mztkll(z') = \mkconst$.
The thin solid curve in Figure~\ref{fig:mock} shows the resulting
flux model for \teff\ for a best-fit value
$\mkconst = 0.028 \; h_{72} \, \rm Mpc^{-1}$.  
The corresponding mean free path $\mmfp = 1/\mkconst = 35.2 \, \umfp$
is in excellent agreement with the traditional
evaluation. 
We stress that the analysis was performed without any consideration of the quasar
SEDs nor any consideration of evolution in the Lyman series line-opacity. 
Although these contribute to the observed flux in the stacked spectrum,
the exponential drop due to \teff\ dominates over this
and any other astrophysical aspect. 

\begin{figure*}
\begin{center}
\includegraphics[height=6.8in,angle=90]{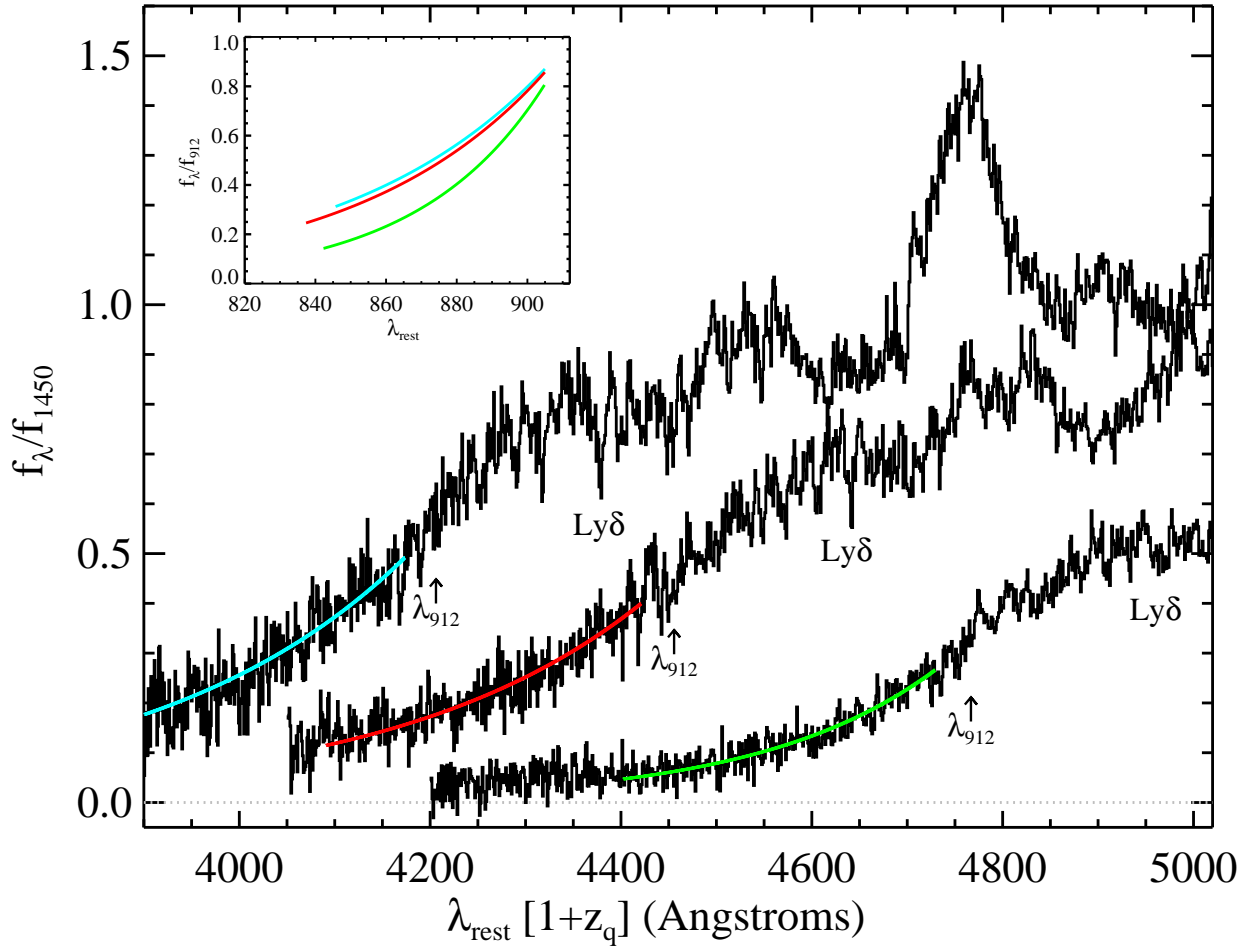}
\end{center}
\caption{The stacked spectrum for three of our redshift bins 
(cyan: $z=\lbrack 3.59,3.63]$; red: $z=\lbrack 3.86,3.92]$; green: 
$z=\lbrack 4.13,4.34]$)
plotted against rest-frame wavelengths redshifted to the mean
quasar redshift for the bin.
Overplotted on each spectrum is our best-fit model
for the absorbed flux below \lll\ due to Lyman limit opacity.
These same curves are shown in the sub-panel against rest wavelengths.
The emission lines are from Lyman series
and metal-line transitions.
}
\label{fig:stacks}
\end{figure*}

\section{Results}

Our new approach provides a tight constraint on the effective mean free path
near the quasar emission redshift (at $z \approx \mzq$).
Because the stacked spectrum covers only several tenths in redshift below
the Lyman limit, it imposes a very weak constraint on the redshift
evolution of \kll.  Instead, one must evaluate
the stacked spectrum of quasars at a range of emission redshifts.

We apply the methodology to 1800 quasar spectra drawn from the SDSS-DR7.
We began with the vetted quasar list from our survey of Lyman limit
systems \citep{pow09} which avoids all purported 
quasars in the SDSS-DR7 that have erroneous redshift estimates
or are not bona-fide quasars.
We have also ignored all quasars with strong associated absorption
in the \ion{C}{4}, \ion{N}{5}, and/or \ion{O}{6} doublets
(e.g.\ broad absorption line systems).  We have not, however, removed quasars
with evident Lyman limit absorption at $z \approx \mzq$.  It is our
goal to estimate the entire Lyman limit opacity that quasars
experience, except for the influence of gas on parsec scales.
Therefore, our estimates of \kll\ include opacity from
the quasar's local galactic environment, i.e.\ its proximity region.


The sample was further limited to the following criteria:
(1) $\mzq \ge 3.5$ to insure significant coverage of the Lyman limit
in the SDSS spectra 
and minimize the likelihood that LLS bias the quasar target probability
(but see below);
(2) S/N~$\ge 4$ at $\mlbr = 1450$\AA;
(3) $\mzq < 4.35$ to insure that a stack of 150 quasars covers a
redshift interval $\Delta z < 0.4$.
Starting at $z=3.5$, we constructed a series of bins 
of 150 quasars each to produce a stacked
spectrum.
Each quasar spectrum was normalized by the observed flux at 
$\mlbr = 1450$\AA\ (in a 20\AA\ window) and shifted to its rest-frame
(nearest pixel).  The full ensemble was then averaged (without
weighting) ignoring bad pixels.  A sample of three of the stacked
spectra are given in Figure~\ref{fig:stacks}. 


As noted above, our analysis of \teff\ includes contributions
from the quasar's proximity region.  In general, this corresponds
to $\approx 10\,\rm Mpc$ or $\Delta \mlbr \approx 9$\AA\
\citep[e.g.][]{dww08}.  From the stacked spectrum, 
Figure~\ref{fig:stacks} shows that the flux at the `edge' of the
proximity region (i.e.\ at $\mlbr \approx 900$\AA) is
$\approx 0.75$ times the flux at 912\AA, giving 
$\mteff \approx 0.3$.  In the absence of the IGM beyond the proximity
region, the flux would begin to recover shortward of 900\AA\ as
$\nu^{-3}$.  It is evident, therefore, that opacity from the `true'
IGM dominates our analysis.  Furthermore, we find that our analysis yields
models that extrapolate well to 912\AA\ suggesting the opacity of
the proximity region follows the behavior of the general IGM.
This is consistent with our finding 
that there is no strong differences in the
incidence of LLS near quasars \citep{pow09}.

Using the model for \teff\ (Equation~\ref{eqn:newteff}),
we have fitted the data to evaluate the opacity \kconst\ at
$z \approx \mzq$ in a series
of redshift intervals, each containing 150 quasars.
We minimized $\chi^2$ over rest-wavelength intervals
starting at 905\AA\ (to minimize bias from strong Lyman series absorption
in the proximity zone of the quasar) and 
extending down in wavelength corresponding to the larger of 3900\AA\ 
and $\mzq-z_{912} = 0.4$.   The scatter in the stacked spectrum
was estimated locally in 21~pixel bins centered at each data point
and presumed to be Gaussian.  
The flux at \lll\ was estimated from the data but allowed to
vary by 10\%\ when minimizing $\chi^2$.  
Because the observed scatter in the stacked spectrum is systematic
(related to stochasticity in the IGM) and is not included in our
model, one cannot estimate $\sigma(\mkconst)$
the uncertainty in \kconst\ from standard
$\chi^2$ techniques.  Instead, we performed a bootstrap
analysis of 100 realizations of each stacked spectrum and estimated
$\sigma(\mkconst)$ from the resulting distribution of \kconst\ values.

\begin{figure*}
\begin{center}
\includegraphics[height=6.8in,angle=90]{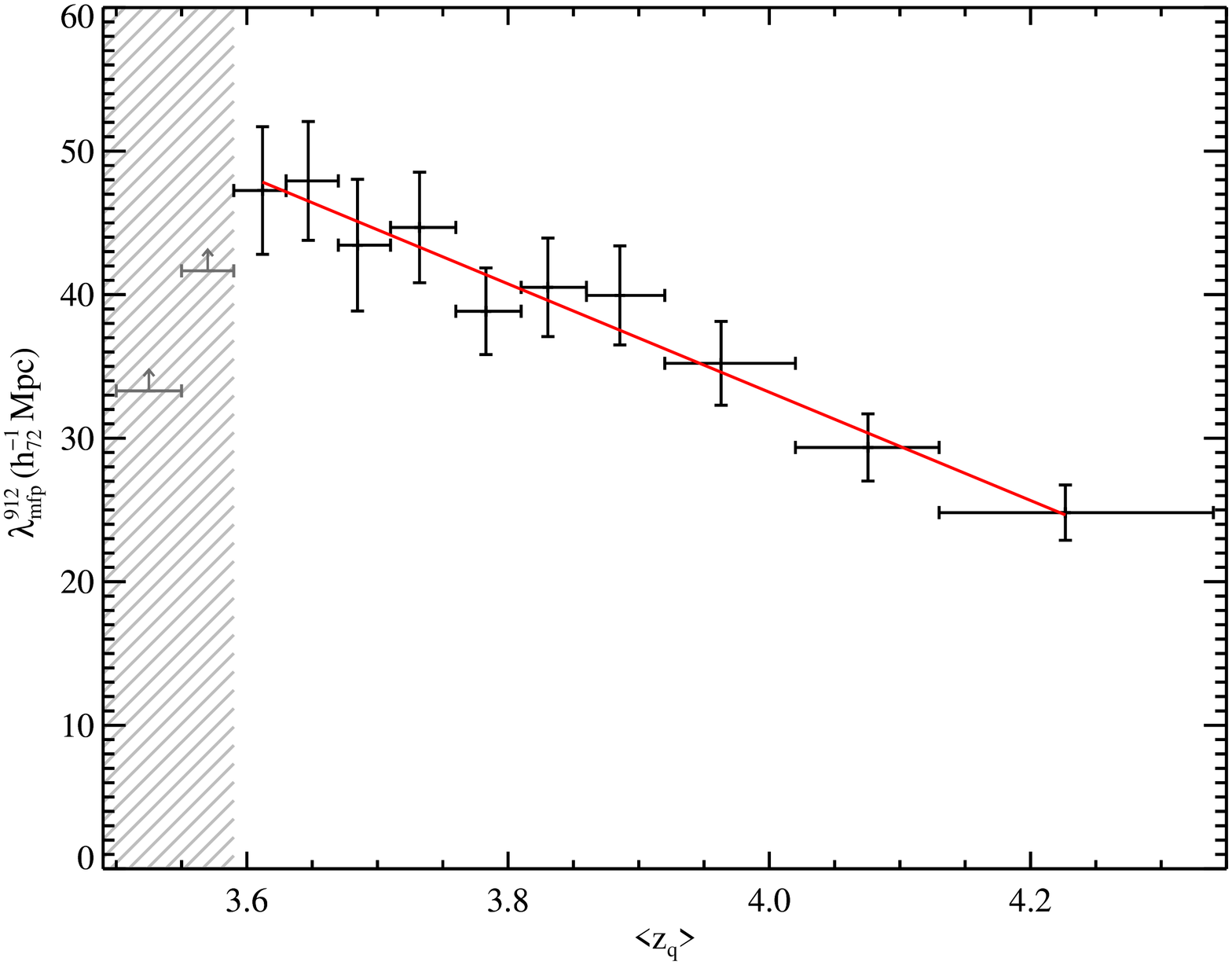}
\end{center}
\caption{Estimates of the mean free path \mfp$\equiv (1/\mkconst)$ 
versus redshift.
Error bars were estimated from standard bootstrap analysis.
The red curve is a simple linear regression of the binned data:
$\mmfp = \bestmfp - \slopemfp (z-3.6)$ with 
$\bestmfp = (\vbestmfp) \, \umfp$
and $\slopemfp = (\vslopemfp) \, \umfp$.
These are the most precise estimates of \mfp\ to date
and provide the first robust description of its redshift
evolution.
The values at $z<3.6$ (shaded region)
are severely affected by a systematic
bias in the SDSS database (see text) and are presented here
only as lower limits.
}
\label{fig:mfp}
\end{figure*}

Figure~\ref{fig:mfp} presents the evaluations of \kconst\
in terms of the mean free path ($\mmfp =1/\mkconst$).
The mean free path exhibits a peak (minimum in opacity)
of nearly $50 \, \umfp$ at $z=3.6$ and 
declines with increasing redshift.  We have parameterized
the redshift evolution in \mfp\ with a simple
linear regression: $\mmfp = \bestmfp - \slopemfp(\mzq-3.6)$.
Restricting the analysis to $z \ge 3.59$, a $\chi^2$ minimization
of this model to the binned evaluations of \mfp\ gives
$\bestmfp = (\vbestmfp) \, \umfp$ and 
$\slopemfp = (\vslopemfp) \, \umfp$.  

\begin{figure*}
\begin{center}
\includegraphics[height=6.8in,angle=90]{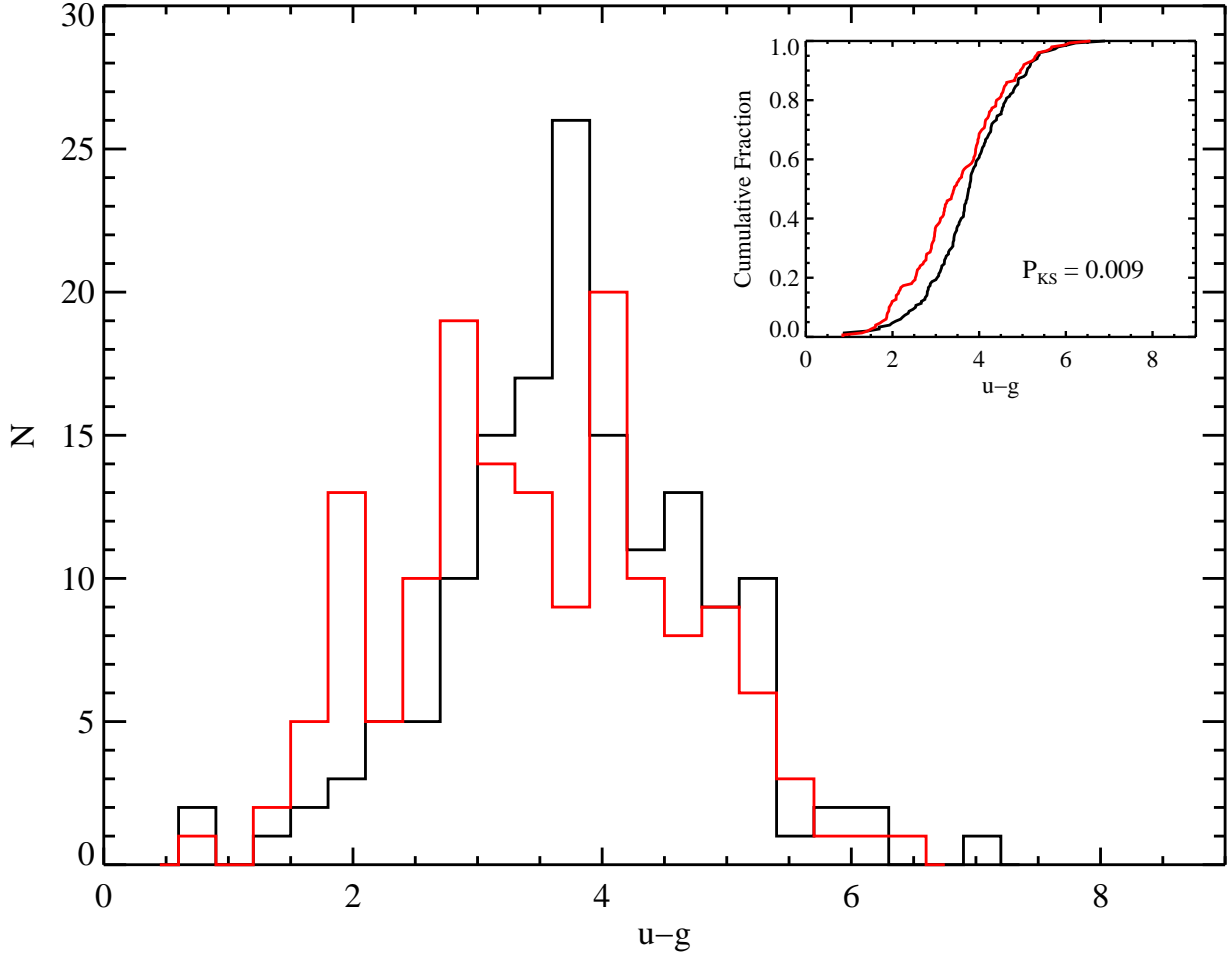}
\end{center}
\caption{Histograms of the \ug\ colors of quasars in the
redshift intervals $z=\lbrack 3.50,3.55]$ (black) 
and $z=\lbrack 3.59,3.63]$ (red).  The lower redshift cohort
shows systematically redder \ug\ colors and a two-sided 
Kolmogorov-Smirnov test rules out the null hypothesis at greater
than $99\%$ c.l.
This offset in \ug\ color contradicts standard 
expectation of a universe where \ion{H}{1}
absorption monotonically increases with redshift.
It occurs because of a bias in the SDSS quasar sample
related to the survey's targeting criteria \citep{rfn+02}.
Note, objects with $\mug \gtrsim 4.5$ are generally not
detected in the $u$-band.
}
\label{fig:ug}
\end{figure*}

Formally, our analysis also indicated a rise in the opacity at
$z<3.6$, i.e.\ the lower limits in Figure~\ref{fig:mfp}.
This runs contrary to all expectation and current understanding
of the IGM. 
Initially, we suspected that
this measurement indicated a systematic error in the SDSS
spectra at the bluest wavelengths \citep[e.g.][]{bss+03}.  To test this
hypothesis, we examined the \ug\ colors of the
quasars in the first ($z \approx 3.5$) and third ($z \approx 3.6$)
quasar bins.
Figure~\ref{fig:ug} histograms the two \ug\ distributions.
The colors of the $z=3.5$ quasars are systematically
{\it redder} than those at $z=3.6$; 
this is the exact opposite of what one predicts if the
IGM were monotonically increasing in opacity with redshift.
We conclude that the quasars at $z=3.5$ drawn from the SDSS
are redder than the cohort at $z=3.6$ because of an elevated
incidence of Lyman limit opacity, 
confirmed by analysis on the incidence of LLS \citep{pow09}.

We then explored whether this elevated opacity is related
to observational bias in the SDSS quasar sample.
We simulated the SDSS experiment by constructing
mock quasar spectra at $z \approx 3.5$ and $z \approx 3.6$
with intrinsic SEDs having mean \ug\ color of 0.57\,mag and
standard deviation of 0.19\,mag.
These spectra were blanketed with IGM absorption assuming a 
monotonically increasing opacity with redshift.
After restricting the quasar sample according to the SDSS
color-selection criteria \citep{rfn+02}, 
we found the $z \approx 3.5$ cohort has
systematically redder colors than the higher redshift sample.
Similarly, we find
a correspondingly higher opacity inferred from the stacked
spectrum.    Because of the targeting criteria,
the cohort of $z \approx 3-3.6$ quasars in the SDSS spectroscopic
database are systematically biased against having 
$\mug < 1.5$ which
biases against sightlines without strong Lyman limit absorption.
Our analysis indicates that the bias extends to $\mzq = 3.6$, beyond
which very few quasars are predicted to ever have such blue color
\citep[see][for further details]{wp09}.
The results for $z>3.6$ are presented in Table~\ref{tab:summ}.

\begin{deluxetable}{lccccccc}
\tablewidth{0pc}
\tablecaption{SUMMARY TABLE\label{tab:summ}}
\tabletypesize{\footnotesize}
\tablehead{\colhead{z} & \colhead{$<\mzq>$} & \colhead{$\lambda_{\rm analysis}$} & \colhead{$\mkconst$} & \colhead{$\sigma(\mkconst)$} \\ 
&& (\AA)& (Mpc$^{-1}$) & (Mpc$^{-1}$) }
\startdata
$\lbrack$3.59,3.63]&3.61&$\lbrack$846, 905]&0.0209&0.0023\\
$\lbrack$3.63,3.67]&3.65&$\lbrack$839, 905]&0.0218&0.0021\\
$\lbrack$3.67,3.71]&3.68&$\lbrack$834, 905]&0.0224&0.0026\\
$\lbrack$3.71,3.76]&3.73&$\lbrack$835, 905]&0.0212&0.0020\\
$\lbrack$3.76,3.81]&3.78&$\lbrack$836, 905]&0.0257&0.0020\\
$\lbrack$3.81,3.86]&3.83&$\lbrack$837, 905]&0.0254&0.0024\\
$\lbrack$3.86,3.92]&3.89&$\lbrack$837, 905]&0.0243&0.0024\\
$\lbrack$3.92,4.02]&3.96&$\lbrack$839, 905]&0.0292&0.0026\\
$\lbrack$4.02,4.13]&4.08&$\lbrack$840, 905]&0.0341&0.0033\\
$\lbrack$4.13,4.34]&4.23&$\lbrack$842, 905]&0.0403&0.0036\\
\enddata
\tablecomments{Because of the color-criteria bias discussed in the paper, we caution that the values for the first few bins may systematically underestimate 
\kconst\ by $10-30\%$.  
All wavelengths are in the quasar rest-frame and all distances are proper.  The assumed cosmology has $\omegam = 0.7, \;\Omega_\Lambda = 0.7, \; {\rm and} \; H_0 = 72 \, \mkms \rm Mpc^{-1}$.}
 
\end{deluxetable}

\section{Discussion}
\label{sec:discuss}

Our new technique provides the most precise measurements on
the IGM opacity to \ion{H}{1} ionizing radiation at any redshift.
Previous estimates of \kll\ were limited by large uncertainties
in the \nhi\ frequency distribution, especially the incidence of
systems with $\mnhi \approx 10^{17} \cm{-2}$ 
\citep{mhr99,sb03,me03,mw04,flh+08}.  Our results indicate
that systems dominating \kll\ at $z>3.6$ must evolve
more slowly than the $\mnhi \le 10^{14} \cm{-2}$ \lya\ forest.
In turn, we infer a flattening in the \ion{H}{1}
frequency distribution between $\mnhi \approx 10^{15}-10^{17} \cm{-2}$ 
\citep{petit93,opb+07}.
We provide a full discussion on the implications for \fnhi\ in \cite{pow09}.

Our analysis also gives the first direct description of the 
evolution in \mfp, albeit over a small redshift
interval.  Our results are well parameterized by a linear
decrease in \mfp\ but can also be described by a $(1+z)^{-\gamma}$
power-law with $\gamma = 3.5 - 5.5$.  These values are consistent
with the observed evolution in the incidence of LLS \citep{pow09}.

Our results refine recent inferences 
\citep[e.g.][]{flh+08,dww09}
that galaxies contribute significantly
to the EUVB at $z>3$.
We assume the photoionization rate at $z=4$ inferred from 
the effective \lya\ opacity of the IGM 
\citep[$\log\mgigm = -12.3$;][]{fpl+08}.  Comparing this
value against the photoionization rate
inferred from quasars \gq\ using an emissivity
$\meq = 2\sci{24} \; {\rm erg s^{-1} Hz^{-1} Mpc^{-3}}$
\citep[comoving;][]{hoprich07,bong07} 
and adopting our estimate of \mfp\ at $z=4$, 
we find $\mgq = 0.5 \, \mgigm$.
This suggests a modest but non-negligible contribution
from galaxies to the EUVB at this redshift.  The systematic
uncertainties
in $\meq$ and \gigm\ are sufficiently large that one could
recover \gigm=\gq, but 
\gq\ would overpredict \gigm\ at $z=3.5$
given the observed rise in \mfp\ and $\meq$ with 
decreasing redshift.

The results also revise at least some previous estimates of the EUVB.
\cite{hm96} and their subsequent analyses (CUBA), for example, 
have adopted an approximately 3 times
shorter mean free path at $z \sim 4$ than our analysis reveals.
This implies: 
(1) the normalization of their
EUVB spectrum is several times too low;
and
(2) the EUVB spectrum is softer at energies of $\approx 1$,\,Ryd.  
The latter point may help to reconcile apparent
contradictions in the metal-line analysis of the IGM \citep{ads+08}
without resorting to a large input from galaxies.
Other analyses, however, have used estimates
for \mfp\ that are in much better
agreement with our results \citep[e.g.][]{flz+09}.
An understanding of the full implications of our new 
constraints on \mfp\ awaits new calculations of the EUVB.


We identified a previously unreported
systematic bias in the SDSS quasar spectroscopic sample
against sightlines at $z < 3.6$ that are `clear'
of optically thick absorbers.
This bias has implications for a range of IGM analysis
at $z \sim 3$ including:
(i) the paucity of sightlines for studying \ion{He}{2} reionization \citep{wp09};
(ii) an overestimate of the incidence of LLS \citep{pow09} and
damped \lya\ systems \citep{pw09};
and
(iii) analysis of the \lya\ forest.
We caution that all existing studies of the IGM
at $z \sim 3$ using the SDSS database
should be reviewed in light of this systematic bias.

In future work, 
we analyze these same spectra to constrain the SEDs
of $z > 3.5$ quasars, infer the \nhi\ frequency distribution 
and Doppler parameters of gas with $\mnhi \approx 10^{16} \cm{-2}$, 
and isolate the opacity of the IGM far from the quasar's 
proximity region.
The technique introduced here is easily extended to higher and lower
redshifts by obtaining modest signal-to-noise (S/N), 
low-resolution spectroscopy of several hundred quasars.
Future ground and space-based programs will precisely estimate
\kll\ from $\mzq \approx 0.5 - 5$.

\acknowledgments

We acknowledge valuable conversations with P. Madau, 
J. Hennawi, G. Richards, and S. Burles.
J. X. P. and J.M.O are supported by NASA grant
HST-GO-10878.05-A.  J.X.P. and G.W. acknowledge
support from NSF CAREER grant (AST--0548180) and
NSF grant AST-0908910.

%


\end{document}